# Current percolation model for the special resistivity behavior observed in Cu-doped Apatite

Qiang Hou[1], Wei Wei[1], Xin Zhou[1], Xinyue Wang[1], Yue Sun[1*], ZhiXiang Shi[1*]

[1] Key Laboratory of Quantum Materials and Devices of Ministry of Education,
School of Physics, Southeast University, Nanjing 211189, China

**Abstract**: Since the initial report of the potential occurrence of room-temperature superconductivity under normal pressure [arXiv: 2307.12008], there has been significant interest in the field of condensed matter physics regarding Cu-doped Apatite ($Pb_{10-x}Cu_x(PO_4)_6O$). In this study, we performed temperature-dependent resistivity measurements on the synthesized $Pb_{10-x}Cu_x(PO_4)_6O$ samples. The structure of the sample was confirmed to match the reference literature through X-ray diffraction analysis. Remarkably, we observed four distinct types of resistivity behaviors within samples from the same pellet: (1) A semiconductor-like behavior characterized by a decrease in resistivity as temperature is lowered. (2) A gradual reduction in resistivity, reaching an exceptionally small value that falls below the resolution limits of our measurement equipment. (3) An abrupt drop in resistivity to a low value at ~ 250 K. (4) An almost linear reduction in resistivity exhibiting a transition at approximately 7 K (possibly associated with Pb). Following a thorough compositional analysis, we proposed a current percolation model, based on the formation of a Cu/Pb current channel, to elucidate the observed special resistivity behaviors. It is important to note that the Meissner effect was not observed in our magnetization measurements. Consequently, we reached the conclusion that the presence of superconductivity in Cu-doped Apatite has yet to be substantiated.

**Introduction**

Room temperature superconductivity, a remarkable scientific feat, is set to enable materials to conduct electricity with absolute zero resistance at ordinary temperatures. This groundbreaking discovery will have immense potential for enhancing energy efficiency and driving technological innovation across various industries, including electronics, transportation, and beyond [1,2,3].

Recently, Lee et al. reported a announcement about the discovery of a room-temperature superconductor under atmospheric pressure, $Pb_{10-x}Cu_x(PO_4)_6O$ with $0.9 < x < 1.1$ [4,5]. They substantiated the presence of room temperature superconductivity in $Pb_{10-x}Cu_x(PO_4)_6O$ through three lines of evidence: (1) a sudden resistivity drop in electrical measurements to near zero, (2) the observation of diamagnetic signals and magnetic levitation phenomena in magnetic measurements, and (3) the sudden voltage jump in measurements of applied current versus voltage.

This revelation instantaneously captured the interest of a significant number of researchers. At present, however, there is no definitive experimental confirmation of room temperature superconductivity. The experiment yields the following different findings: (1) absence of both zero resistance and magnetic levitation, (2) presence of magnetic levitation without zero resistance [6,7], and (3) presence of nearly zero resistance without magnetic levitation characteristics [8]. First of all, two critical properties of superconductors are the Meissner effect and zero resistance. It's essential to emphasize that magnetic levitation is not synonymous with the Meissner effect. In terms of verifying magnetic levitation phenomenon, the majority of researchers have already observed this phenomenon in their experiments. Kaizhen Guo et al. also observed "half levitation" and found that the samples ubiquitously contain weak yet definitive soft ferromagnetic components [7]. They proposed that, in conjunction with the distinct shape anisotropy of the small fragments, the presence of soft ferromagnetism adequately accounts for the observed phenomenon of half levitation within intense vertical magnetic fields. During the experimental verification of zero resistance, Li Liu et al. [9] observed that as the temperature decreases, the resistivity of LK-99 increases. This observation suggests that LK-99 exhibits semiconductor or insulator characteristics. Shilin Zhu et al. [10] observed $Cu_2S$ impurities in the reported LK-99 sample. They examined the transport and magnetic traits of pure $Cu_2S$ and $Cu_2S$-inclusive LK-99. They argued that the so-called superconducting behavior in LK-99 is most likely due to a reduction in resistivity caused by the first order structural phase transition of $Cu_2S$ at around 385 K, from the $\beta$ phase at high temperature to the $\gamma$ phase at low temperature. A magnitude reduction in resistivity caused by the first-order structural phase transition of $Cu_2S$ might clarify Lee et al.'s results [4,5]. However, the significantly higher resistivity of $Cu_2S$ contradicts the extremely small resistivity observed below 110K in our prior study [8]. Furthermore, our sample had a lower presence of $Cu_2S$ impurities.

As for theory, multiple independent research groups [11,12,13,14,15] conducted Density Functional Theory (DFT) [16] calculations, yielding similar findings: Within the lead apatite crystal structure, when one Pb atom is substituted by Cu, it induces exceptionally flat bands near the Fermi level. This phenomenon is analogous to what has been observed in twisted bilayer graphene [17,18], where the presence of flat bands close to the Fermi level signifies the potential for superconductivity. Therefore, some researchers propose that the flat DFT bands crossing the Fermi energy indicate that Cu doping $x \approx 1$ transforms the insulating lead apatite into a metallic state, thus explaining the conducting and possibly superconducting nature of LK-99 [11,12,14,15]. Held et al. [13] provided *ab initio* derived tight-binding parameters for a two- and five-band model. Their research implies that the interaction-to-bandwidth ratio categorizes LK-99 as either a Mott or charge transfer insulator. Achieving metallic and potentially superconducting behavior would necessitate electron or hole doping.

In this report, we achieved successful synthesis of the $Pb_{10-x}Cu_x(PO_4)_6O$ compound and observed four distinct resistivity behaviors within the same sample. We proposed a current percolation model, based on the formation of a Cu/Pb current channel, to elucidate the observed special resistivity behaviors.

**Experiment**

Detail about the sample growth can been seen in our previous report [8]. The crystal structure of produced samples was examined using an X-ray diffractometer (XRD) with Cu-Kα radiation on a Bruker D8 Discover diffractometer. Elemental analysis was performed by a scanning electron microscope equipped with an energy dispersive x-ray spectroscopy (EDS) probe. The in-plane electrical resistivity was carried out by the standard four-probe method on a physical property measurement system (PPMS, Quantum Design). Magnetization measurement was preformed by the VSM (vibrating sample magnetometer) option of PPMS.

**Results and Discussion**

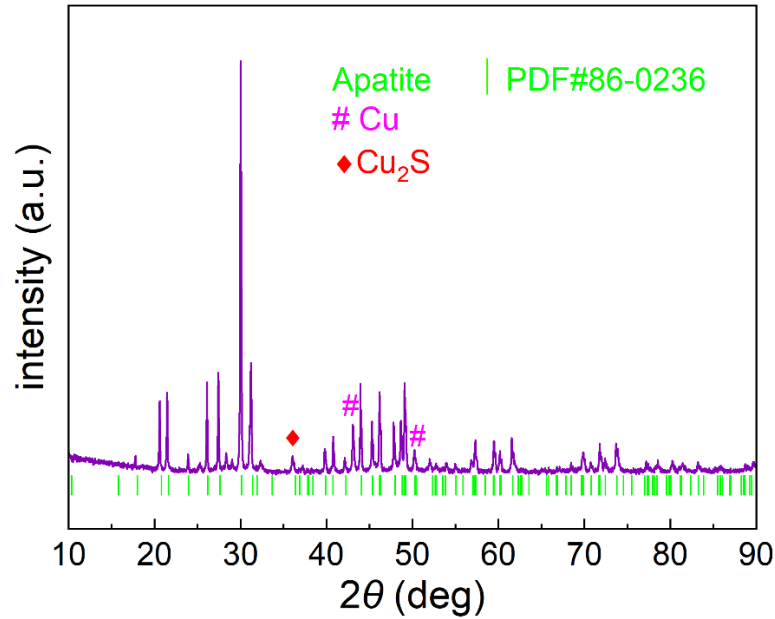

Fig. 1 X-ray diffraction pattern of sample, with the main phase $Pb_{10-x}Cu_x(PO_4)_6O$ and impurity phases of Cu and $Cu_2S$.

To validate the successful synthesis of $Pb_{10-x}Cu_x(PO_4)_6O$, Figure 1 displays the X-ray diffraction (XRD) pattern of the sample, with diffraction peaks consistent with those in the ref. [4].

Composition analysis shows that in addition to the main phase $Pb_{10-x}Cu_x(PO_4)_6O$, sample also contains some amounts of Cu and Cu-S compounds, which is highly consistent with many recent reports [6,7,9,19] and provides us clue for exploring the nature of many phenomena akin to superconductivity in $Pb_{10-x}Cu_x(PO_4)_6O$. It is worth mentioning that our sample has better phase formation compared to other teams according to the XRD, and the composition of the impurities is different, which may be one of the reasons for the different resistivity results we observed.

Figure 2(a) illustrates the specimen we prepared, meticulously fashioned into a rectangular form with precise measurements: a length of 3mm, a width of 1mm, and a thickness of 0.5mm. Through an analysis of Energy Dispersive Spectroscopy (EDS) on the whole sample (see Fig. 2(c)), the elemental composition ratio of Pb:Cu:P was determined to be approximately 1.51:2.27:1. This ratio establishes a resemblance to the lead and phosphorus proportions found in the previous report [4,5]. However, it is important to highlight the considerably elevated Cu content present within our sample. In reference to Fig. 2(b), it becomes evident that the sample harbors a substantial amount of Cu particles. This pronounced abundance is attributed to an excessive precipitation of Cu during the preparation process. As a result, the localized EDS analysis, depicted in Fig. 2(d) within the red-circled area in Fig. 2(b), discloses a remarkably high Cu content of 92%. To provide further insight into the Cu distribution, the mapping of lead, copper, and phosphorus derived from EDS surface scans are shown in Fig. 2(e), (f), and (g) respectively. Obviously. The Cu is inhomogeneously distributed, and some islands composed by copper (and small amount of Pb) can be identified.

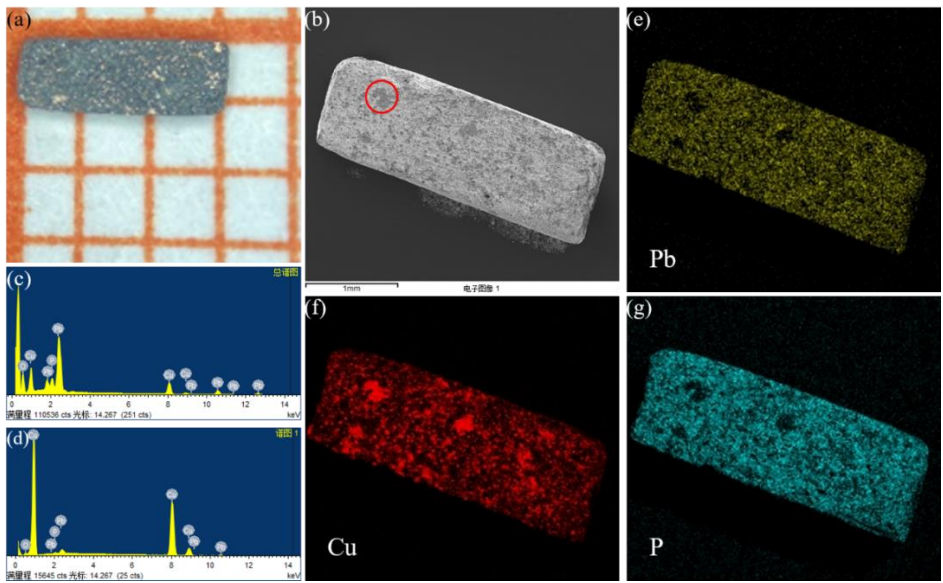

Fig. 2: (a) A photo of the sample. (b) A scanning electron microscope (SEM) image of sample. The Energy-dispersive X-ray spectroscopy (EDS) of (c) the whole sample, (d) and the region

circled in red in (b). The Energy-dispersive X-ray spectroscopy mapping of (e) lead, (f) copper, and (g) phosphorus.

In order to further investigate the physical properties of $Pb_{10-x}Cu_x(PO_4)_6O$, we conducted electrical transport and magnetization measurements. Figs. 3(a-d) depict four distinct *R-T* (resistivity-temperature) curves acquired from various regions of the same sample. Fig. 3(a) reveals a semiconducting behavior, resembling recent findings [9,19]. Interestingly, Fig. 3(b) displays metallic behavior, and an extremely small resistivity emerges below 110 K, as more clearly illustrated in the inset of Fig. 3(b). Data shown in Fig. 3(b) is the same as that presented in our previous report [8], but in the linear plot. In Fig. 3(c), an abrupt drop in resistivity to a small value at ~ 250 K was observed. The enlarge of the low temperature part can be seen in the inset of Fig. 3(c). We also measured the temperature dependence of resistivity under fields up to to 9 T. As the applied magnetic field increasing, the resistivity drop is gradually suppressed. In Fig. 3(d), resistivity manifests a good linear behavior and a sharp transition at 7.1 K, strongly indicating the presence of Pb although it is not detected via XRD.

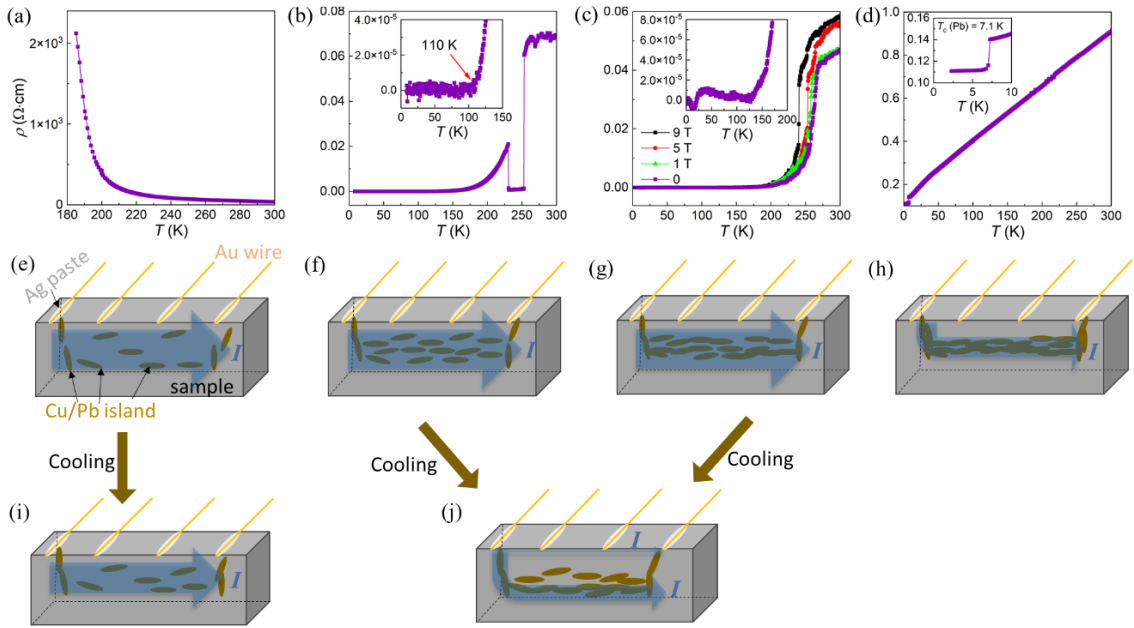

Fig. 3 Temperature dependence of resistivity obtained from four measurements on the different regions of the same sample (a) semiconducting, (b) extremely small resistivity below 110 K, (c) abrupt resistivity drop at ~ 250 K, (d) linear behavior with transition at 7.1 K. (e-j) Current-percolation model. The brown ellipses depict the Cu/Pb islands, while the blue arrow illustrates the current flow.

To comprehensively explain the observed variety of resistivity behaviors, we have put forth a current-percolation model, rooted in the formation of Cu/Pb islands within the sample. In most instances, these Cu/Pb islands remain isolated across the entire temperature range, as illustrated in Figs. 3(e) and 3(i). Consequently, the current traverses through the entirety of the sample, navigating both the Cu/Pb islands and the $Pb_{10-x}Cu_x(PO_4)_6O$. This results in a discernible semiconducting behavior, which was observed in a majority of our samples and has been reported by other research groups as well [9,19].

In specific regions of the sample, the concentration of Cu/Pb islands may be higher, exemplified in Fig. 3(f). As the temperature decreases, these initially isolated Cu/Pb islands approach each other due to potential volume contraction. In this context, the resistivity experiences a gradual reduction because the path of the current incorporates more of these low-resistance Cu/Pb islands. Ultimately, these islands can coalesce to form a continuous channel, which could manifest as a chain-like, net-like, or even more intricate structural configuration. For the sake of simplicity, we've depicted a Cu/Pb chain to represent this channel, as depicted in Fig. 3(j). In this scenario, a significant proportion of the current traverses through the low-resistance Cu/Pb channel, with only a negligible amount passing through the surface regions, subsequently being detected by the contacts. This results in an exceedingly minimal voltage drop, leading to the recording of an extremely low resistance or even a resistance akin to "zero" when the voltage drop is smaller than the resolution limit of the measuring equipment. This model effectively elucidates the resistivity behavior observed in Fig. 3(b).

In certain regions, a substantial portion of Cu/Pb islands are interconnected even at room temperature, although the formation of a continuous channel is impeded by specific breaking points, as depicted in Fig. 3(g). As the temperature decreases, these breaking points gradually connect, leading to the establishment of a Cu/Pb channel, as illustrated in Fig. 3(j). Consequently, the resistivity experiences an abrupt and substantial decrease to a remarkably low value, explaining the observations in Fig. 3(c). Moreover, the application of magnetic fields enhances the conductivity path, making the process of Cu/Pb channel formation more challenging. Consequently, the point at which the resistivity dramatically drops shifts to lower temperatures under magnetic fields, as observed in the Fig. 3(c).

In an extreme scenario, the Cu/Pb channel formation is already accomplished at room temperature, as illustrated in Fig. 3(h). In this specific instance, the resistivity measurement reflects that of the Cu/Pb material. Notably, the superconducting transition originating from Pb at approximately 7.1 K has also been detected, as indicated in the inset of Fig. 3(d).

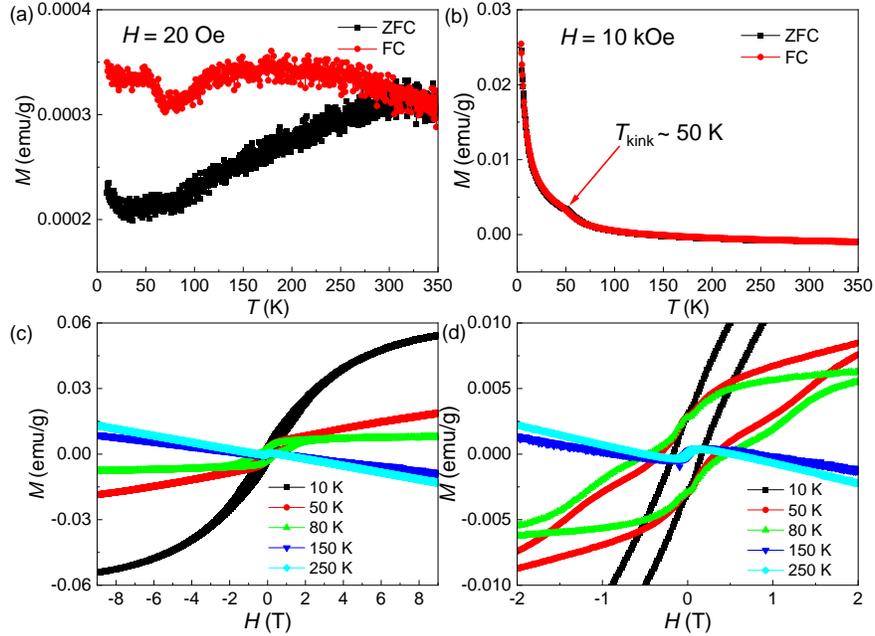

Fig. 4 The magnetization of sample (a) The magnetization versus temperature (*M-T*) at a external magnetic field of 20 Oe, (b) *M-T* at an external magnetic field of 10 kOe, (c) *M-H* loop at different temperatures, (d) enlarge view of *M-H* loop at the low field.

The magnetization measurements results are illustrated in Fig. 4. In Fig. 4(a), both the field cooling (FC) and zero-field-cooling (ZFC) exhibit positive magnetic moments. A significant branching between ZFC and FC, and an abnormality at 100 K was also observed. When increase magnetic field to 10 kOe, the FC and ZFC curves nearly coincide, erasing the branching effect, as depicted in Fig. 4(b). As the temperature decreases, the magnetization value changes from negative to positive around 125 K, while the ZFC curve displays a kink at 50 K, which may be due to an order state from the impurities.

We also conducted *M-H* loop measurements at varying temperatures. In Fig. 4(c), the *M-H* curve at high field and low temperatures displays a positive slope, indicating prevailing paramagnetism. Conversely, at high temperatures, 150 K and 250 K, a negative slope manifests, pointing to prevailing antimagnetism. Zooming in on the *M-H* loop at low fields, as shown in Fig. 4(d), a distinct magnetic hysteresis loop emerges, spanning the entire temperature range from 10 K to 250 K. This unequivocally confirms the presence of ferromagnetic ordering. It is noteworthy that the *M-H* loops at 10 K, 50 K and 80 K exhibit kinks near 0 T, which need future analysis. More importantly, no Meissner signal or superconducting-like *M-H* loop was observed.

**Conclusion**

In summary, we have introduced a current percolation model as an explanatory framework for the diverse resistivity behaviors observed in Cu-doped Apatite. Given the absence of a Meissner signal, we draw the conclusion that there is still no evidence for the superconductivity in Cu-doped Apatite.

**Acknowledge**

We thank the valuable comments, suggestions and understanding from many scientists most of them are from China. This work was partly supported by the National Key R&D Program of China (Grant No. 2018YFA0704300), the Strategic Priority Research Program (B) of the Chinese Academy of Sciences (Grant No. XDB25000000), and the National Natural Science Foundation of China (Grant No. U1932217).

These authors contributed equally: Qiang Hou, Wei Wei, Xin Zhou.
*Corresponding author: sunyue@seu.edu.cn;
*Corresponding author: zxshi@seu.edu.cn.